\documentclass{article}
\usepackage{natbib}
\bibpunct{(}{)}{;}{a}{}{,}
\usepackage{emulateapj}
\usepackage{epsfig}
\newcommand{\clb}{{\mathcal{B}}}
\newcommand{\cle}{{\mathcal{E}}}
\newcommand{\clp}{{\mathcal{P}}}
\newcommand{\uD}{\mathrm{D}}
\newcommand{\curl}{\mathrm{curl}}
\newcommand{\CMBFAST}{\textsc{CMBFAST}}
\newcommand{\Omtot}{\Omega_{\mathrm{tot}}}
\newcommand{\Omb}{\Omega_{\mathrm{b}}}

\newcommand{\Omm}{\Omega_{\mathrm{m}}}
\newcommand{\Oml}{\Omega_\Lambda}
\lefthead{Lewis, Challinor, \&\ Lasenby}
\righthead{CMB anisotropies in closed {FRW} models}
\begin{document}
\title{EFFICIENT COMPUTATION OF CMB ANISOTROPIES IN CLOSED FRW MODELS}
\author{Antony Lewis\altaffilmark{1}, Anthony Challinor\altaffilmark{2}, \&\
Anthony Lasenby\altaffilmark{3}}
\altaffiltext{1}{A.M.Lewis@mrao.cam.ac.uk}
\altaffiltext{2}{A.D.Challinor@mrao.cam.ac.uk}
\altaffiltext{3}{A.N.Lasenby@mrao.cam.ac.uk}
\affil{Astrophysics Group, Cavendish Laboratory, Madingley Road, Cambridge CB3 0HE, UK.}
\begin{abstract}
We implement the efficient line of sight method to calculate the anisotropy
and polarization of the cosmic microwave background for scalar
and tensor modes in almost-Friedmann-Robertson-Walker models with positive
spatial curvature. We present new results for the polarization power spectra
in such models.
\end{abstract}
\section{INTRODUCTION}

The anisotropy of the cosmic microwave background (CMB) plays a key role in
many areas of modern cosmology. Joint analyses of current CMB and
type Ia supernovae data~\citep[e.g.][]{Efstathiou99,Turner99} suggest that the
universe is within a factor $\Omtot\equiv \Omm+\Oml = 1\pm 0.2$ of the
critical density required for a flat geometry. Closed models
(in which $\Omtot>1$) therefore account for an important sector of the
possible parameter space. Moreover, maximum likelihood searches require
theoretical predictions over a much larger volume of parameter space to
establish reliable error estimates on the parameters under consideration.
It is therefore vital to have
a fast and accurate method for calculating anisotropies for models at least
within the range $0.4< \Omtot< 1.6$. Previous parameter determinations,
such as~\citet{Efstathiou99}, have had to rely on analytical approximations
for rapid calculations of the CMB power spectrum in the closed region, since
the current state of the art codes, such as the widely-used
\CMBFAST~\citep{Seljak96},
do not yet support closed models.

In this \emph{Letter} we describe a numerical implementation of the
linearized equations of the 1+3 covariant approach to CMB
anisotropies~\citep*{LC-scalcmb,chall99a,chall99b,gebbie99b} in
almost-Friedmann-Robertson-Walker (FRW) models with open, flat, and closed
background geometries. Our code thus allows an efficient exploration of the
full cosmological parameter space. We present new results for the polarization
power spectra from scalar and tensor perturbations in closed models. The
intensity power spectra in closed models have been calculated before:
\citet{White96} integrated the full Boltzmann hierarchy directly for scalar
perturbations, extending the earlier semi-analytic predictions
of~\citet{abbott86}. Also,~\citet*{Allen95} used the semi-analytic
approach (adequate on large scales) to calculate the tensor spectrum.
None of these calculations included the effects of polarization.

The 1+3 covariant formalism provides a physically transparent, exact
(fully non-linear) description of both dynamics and radiative transfer in
general cosmological models~\citep*{ellis98,maartens99}. The full formalism
admits a gauge-invariant linearization about FRW models, resulting in a linear
perturbation theory which is arguably simpler, and more physically
transparent, than other approaches~\citep[e.g.][]{ma95,hu98}. Our
implementation of the 1+3 formalism is based on the field-tested
\CMBFAST\ 
code written by~\citet{Seljak96}. Their code uses a line of sight integration
method to achieve high efficiency without compromising accuracy.

\section{IMPLEMENTATION}

\subsection{Basic formalism}

We employ the 1+3 covariant approach to perturbations in
cosmology~\citep[e.g.][]{ellis98},
in which departures from exact FRW symmetry are described by gauge-invariant
variables derived from physical observables relative to some timelike
4-velocity field $u^a$. The equations of radiative transfer can be recast as
propagation equations along the integral curves of $u^a$ for the multipoles
of the total intensity, $I_{A_\ell}$, and the electric and magnetic components
of the total linear polarization, $\cle_{A_\ell}$ and
$\clb_{A_\ell}$~\citep{chall99a}.
Here the multipoles are projected (relative to $u^a$) symmetric trace free
(PSTF) tensors ($A_\ell$ represents the index string $a_1\dots a_\ell$), which
provide a basis-free alternative to the more common
scalar-valued multipole coefficients in spherical harmonic expansions of the
intensity and polarization~\citep*{kamion97,seljak97}.
We only consider linear polarization here, since circular
polarization is not generated by Thomson scattering. For small departures from
an FRW model, the intensity multipoles evolve
as~\citep*{thorne81,ellis83,LC-scalcmb}
\begin{eqnarray}
&& \dot{I}_{A_\ell} + \frac{4}{3}\Theta I_{A_\ell} + \uD^b I_{bA_\ell}
- \frac{\ell}{(2\ell+1)}\uD_{\langle a_\ell}I_{A_{\ell-1}\rangle}
+ \frac{4}{3} \delta_\ell^1 I A_{a_1} \nonumber \\
&&\mbox{}- \frac{8}{15} \delta_l^2 I \sigma_{a_1 a_2}
= -n_{\mathrm{e}} \sigma_{\mathrm{T}}\Bigl[I_{A_\ell} -\delta_\ell^0 I
- \frac{4}{3} \delta_\ell^1 I v_{a_1} \nonumber \\
&&\mbox{} - \frac{1}{10} \delta_\ell^2(
I_{a_1 a_2} + 6 \cle_{a_1 a_2})\Bigr],
\label{eqa}
\end{eqnarray}
where $\Theta$ is the expansion of $u^a$, $\sigma_{ab}$ is the
shear, and $A_a$ is the acceleration. Here, $\uD^a$ is the totally projected
covariant derivative, an overdot denotes the covariant derivative along $u^a$,
and angle brackets denote the PSTF part of the enclosed
indices. The electron number density is $n_{\mathrm{e}}$ in its rest frame
which has relative velocity $v^a$, and the Thomson
cross section is $\sigma_{\mathrm{T}}$.
For the electric polarization, we have~\citep{chall99a}
\begin{eqnarray}
&& \dot{\cle}_{A_\ell} + \frac{4}{3}\Theta \cle_{A_\ell} +
\frac{(\ell+3)(\ell-1)}{(\ell+1)^2}\uD^b \cle_{b A_\ell} \nonumber \\
&&\mbox{} -\frac{\ell}{(2\ell+1)}
\uD_{\langle a_\ell}\cle_{A_{\ell-1}\rangle}-\frac{2}{(\ell+1)}
\curl\clb_{A_\ell} \nonumber \\
&&\mbox{} = -n_{\mathrm{e}}\sigma_{\mathrm{T}}\left[\cle_{A_\ell}-\frac{1}{10}
\delta_{\ell}^{2} (I_{a_1 a_2} + 6 \cle_{a_1 a_2})\right],
\label{eqb}
\end{eqnarray}
and for the magnetic polarization:
\begin{eqnarray}
&&\dot{\clb}_{A_\ell} + \frac{4}{3}\Theta \clb_{A_\ell} +
\frac{(\ell+3)(\ell-1)}{(\ell+1)^2}\uD^b \clb_{b A_\ell} \nonumber \\
&&\!\!\!\!\!\!
-\frac{\ell}{(2\ell+1)} \uD_{\langle a_\ell} \clb_{A_{\ell-1} \rangle}
+ \frac{2}{(\ell+1)} \curl \cle_{A_\ell} = -n_{\mathrm{e}}
\sigma_{\mathrm{T}} \clb_{A_\ell}.
\label{eqc}
\end{eqnarray}
The evolution of the electric and magnetic multipoles are coupled through
the curl terms, where $\curl \cle_{A_\ell} \equiv \epsilon_{b c \langle a_\ell}
\uD^b \cle_{A_{\ell-1}\rangle}{}^{c}$, with $\epsilon_{abc}$ the
alternating tensor in the rest space of $u^a$.
The multipole equations (\ref{eqa}--\ref{eqc}) hold for a general linear
perturbation around an arbitrary FRW model, and for a general choice of $u^a$.
The equations must be supplemented by the
covariant hydrodynamic and gravitodynamic equations~\citep[e.g.][]{ellis98}
to determine the first-order source terms $\sigma_{ab}$, $A_a$ and $v_a$.
For an alternative approach to polarized radiative transfer in general
FRW geometries, see e.g.~\citet{hu98}.

The dimensionless power spectrum of the intensity anisotropies is defined
by the ensemble average~\citep{gebbie98a,LC-scalcmb}
\begin{equation}
C_\ell^{II} = \frac{4\pi}{(4I)^2}\frac{(2\ell)!}{(-2)^\ell(\ell!)^2}
\langle I_{A_\ell} I^{A_\ell} \rangle.
\label{eqd}
\end{equation}
Power spectra for the electric polarization multipoles, $C_\ell^{\cle\cle}$,
the magnetic multipoles, $C_\ell^{\clb\clb}$, and the cross-correlation between
electric polarization and the intensity, $C_\ell^{I\cle}$, can be defined
analogously~\citep{chall99a}. (To conform with~\citealt{seljak97} we
include a factor of $\surd[(\ell+1)(\ell+2)/\ell(\ell-1)]$ on the
right-hand side of eq.~[\ref{eqd}] for each factor of the polarization.)

We solve the multipole equations (\ref{eqa}--\ref{eqc}) by expanding
the first-order variables in PSTF tensors derived from the appropriate
scalar, vector, and tensor eigenfunctions of the comoving
Laplacian $S^2 \uD^a \uD_a$, where $S$ is the scale factor, with eigenvalue
$k^2$. The different perturbation types decouple at linear order,
with each giving rise to a set of coupled first-order differential
equations. The Boltzmann multipole equations for scalar and tensor modes are
given in detail in~\citet{chall99c}. The equations describing
perturbations in the other matter components and the geometry, which determine
the source terms in the Boltzmann hierarchies, can be found
in~\citet{LC-scalcmb,chall99b,gebbie99b}.
The mode-expanded Boltzmann equations can be solved formally as integrals
along the line of sight. These integral solutions form the basis of
the line of sight algorithm employed by \CMBFAST. The solutions for the
intensity are given in the 1+3 covariant formalism
in~\citet{LC-scalcmb,chall99b,gebbie99b}; solutions for the polarization
are given in~\citet{chall99c}. Equivalent results in the total
angular momentum formalism are given in~\citet{hu98}.
The present day multipoles are obtained by integrating the product of source
functions, which are inexpensive to compute, with
special functions (derived from the hyperspherical Bessel functions)
which result from the projections of the eigenfunctions of the Laplacian onto
directions on the sky.

For scalar modes, define $\nu^2\equiv (k^2+K)/|K|$, where $6K/S^2$ is the
curvature scalar of the spatial sections in the background FRW model.
In  the closed case, the regular scalar eigenfunctions of the comoving
Laplacian are complete for $\nu$ an integer
$\geq 1$~\citep[e.g.][]{tomita82,abbott86}. The mode with
$\nu=1$ does not contribute to the perturbations, while the modes with
$\nu=2$ (which can only represent isocurvature
perturbations;~\citealp{bardeen80}) only
contribute to the CMB dipole. For tensor modes $\nu^2\equiv (k^2+3K)/|K|$,
and the regular eigentensors of the Laplacian are complete with $\nu$
an integer $\geq 3$; explicit PSTF representations are given
in~\citet{chall99b}.

\subsection{Numerical Evaluation}

The implementation strategy of \CMBFAST~\citep{Seljak96} requires only minor
modifications for closed models. The modes are now discretized with
wavenumber $\nu~\ge~3$. A given $\nu$ only contributes to multipoles
with $\ell < \nu$, so, unlike the open case, the Boltzmann hierarchies for a
given $\nu$ truncate at finite $\ell$. For large $\nu$, it is
possible to terminate the hierarchies artificially (taking care to avoid
spurious reflection of power) at some lower $\ell$ without compromising
accuracy in the evaluation of the source functions.
The sources are calculated at approximately logarithmically spaced integer
values of $\nu$ and interpolated as needed. 

In closed models, a given linear scale at last scattering subtends a larger
angle on the sky today than in open or flat models. This geometric effect
shifts power in the CMB spectra to smaller $\ell$. Since \CMBFAST\ only
computes the power spectra at a few values of $\ell$, we adjust the
$\ell$-sampling according to the the curvature to maintain accurate
interpolation in all cases.

For the line of sight integral over sources, we calculate the
hyperspherical Bessel functions by integrating a second order differential
equation, as in the open \CMBFAST\ code~\citep*{zaldarriaga98}
The starting values for the Bessel function and its
derivative are found using a recursive evaluation or, for the cases where
it is accurate and faster, the WKB approximation~\citep{Kosowsky98}.
We assume that the development angle $\chi$ satisfies $\chi < \pi$,
where $\chi=\sqrt{|K|}\eta_0$ and $\eta_0$
is the conformal age of the universe.  This
permits calculations with models up to
$\Omtot \approx 1.7$ for a matter fraction $\Omm\approx 0.4$.
For the case $\chi \approx
\pi$, all lines of sight converge to the antipodal point close to last
scattering, and the power spectrum becomes featureless~\citep{White96}.

It is important to maintain numerical stability in the differential
equations for the dependent variables. This is especially true for
scalar perturbations with isocurvature initial conditions, where a poor
choice of dependent variables can lead to large violations of the Einstein
constraint equations unless the initial conditions are specified to exquisite
accuracy. (This problem is particularly acute in the Newtonian
gauge, e.g.~\citealt{ma95}.) We choose to work in a frame in which $u^a$
coincides with the 4-velocity of the cold dark matter (CDM). For scalar
perturbations, we determine the perturbations to the geometry by
evolving the projected gradient of the 3-curvature on hypersurfaces
orthogonal to $u^a$, which gives good numerical stability for all initial
conditions. (With this choice of dependent variables, our equation set
is equivalent to a subset of the widely-used synchronous equations,
gauge-fixed to the CDM.) We allow for adiabatic and two types of isocurvature
initial conditions for scalar perturbations. Also, we provide the
option of specifying the primordial power spectra in non-parametric
form, which is useful for inverse problems such as initial power spectrum reconstruction.

We have verified our calculations against results
obtained with \CMBFAST\ version 2.4.1 for models supported by the latter
(open and flat). We have also compared our results with a pre-release version
of \CMBFAST\ which Seljak \& Zaldarriaga are developing to support closed
models, with good agreement well into the damping tail.

\section{RESULTS}

In Figure~\ref{Cls} we plot the intensity and polarization power spectra
in $\Lambda$CDM models assuming no reionization. One model is closed
($\Omtot=1.2$, $\Oml=0.8$), while the other is flat
($\Omtot=1$, $\Oml=0.6$).
In both cases we take the matter fraction $\Omm=0.4$,
baryon fraction $\Omb=0.045$, and Hubble's constant
$H_0=65 \mathrm{kms}^{-1}\mathrm{Mpc}^{-1}$. For the scalar modes,
we assume adiabatic initial conditions with a scale-invariant primordial
power spectrum $\clp_{\mathrm{S}}(\nu) = \mathrm{constant}$.
Our conventions generalize those of~\citet{lyth95}, so that
the gradient of the 3-Ricci scalar on comoving hypersurfaces
$\tilde{D}_a \tilde{R}{}^{(3)}$, receives power
\begin{equation}
\langle |\tilde{D}_a \tilde{R}{}^{(3)}|^2 \rangle \propto
\sum_{\nu \geq 3}\left(\frac{k}{S}
\right)^6\left(\frac{\nu^2-4}{\nu^2-1}\right)^2\frac{\nu \clp_{\mathrm{S}}
(\nu)}{(\nu^2-1)},
\end{equation}
from super-Hubble modes. The tildes denote that the quantity
on the left is evaluated in the (energy) frame where the momentum density
$q_a$ vanishes. For the tensor modes, we assume a scale-invariant spectrum
$\clp_{\mathrm{T}}(\nu) = \mathrm{constant}$. Our conventions are such that
the power in the electric part of the Weyl tensor
$E_{ab}$~\citep[e.g.][]{ellis98} from super-Hubble modes is
\begin{equation}
\langle E_{ab}E^{ab} \rangle \propto \sum_{\nu \geq 3}
\left(\frac{k}{S}\right)^4 \frac{(\nu^2-4)}{\nu^2}\left(\frac{\nu^2-1}{\nu^2-3}
\right)^2 \frac{\nu \clp_{\mathrm{T}}(\nu)}{(\nu^2-1)}.
\end{equation}
Note that in closed models, the infra-red divergence in the tensor power
spectrum seen in open models with $\clp_{\mathrm{T}}(\nu) = \mathrm{constant}$
is avoided because of the cutoff at $\nu=3$.

The models in Figure~\ref{Cls} have equal physical densities,
$\Omm H_0^2$ and $\Omb H_0^2$, so the sound
horizon at last scattering and the early-time dynamics are approximately equal
in the two models. On small angular scales, where the polarization and scalar
anisotropies are projections of effects at last scattering, restricting
parameter changes to $\Oml$ and the curvature leads to approximate scaling of
the CMB power spectra: $C_\ell \rightarrow C_{\alpha \ell}$ where $\alpha$
is the ratio of the angular-diameter distances to last scattering
in the original and final models. Approximations of this sort were
used by~\citet{Efstathiou99} to include closed models in their recent
joint analysis of CMB and supernovae data.
On large angular scales (small $\ell$), the approximate scaling is
broken for scalar anisotropies by the (late-time) integrated Sachs-Wolfe
effect~\citep[e.g.][]{hu95}, which is responsible for the enhancement at
low $\ell$ seen in the closed model in
Figure~\ref{Cls}~\citep{abbott86,White96}. The effects of curvature terms
in the primordial power spectrum are also potentially observable
on large scales~\citep{hu98}. The presence of the curvature scale in
the primordial power spectrum, combined with the cutoff at
$\nu=3$, suppresses the intensity quadrupole for tensor modes in the closed
model relative to the flat model in Figure~\ref{Cls} \citep[c.f. the
case of open models, e.g.][]{hu97}. 

\begin{center}
\epsfig{figure=plot.eps,angle=-90,width=8.5cm}
\end{center}
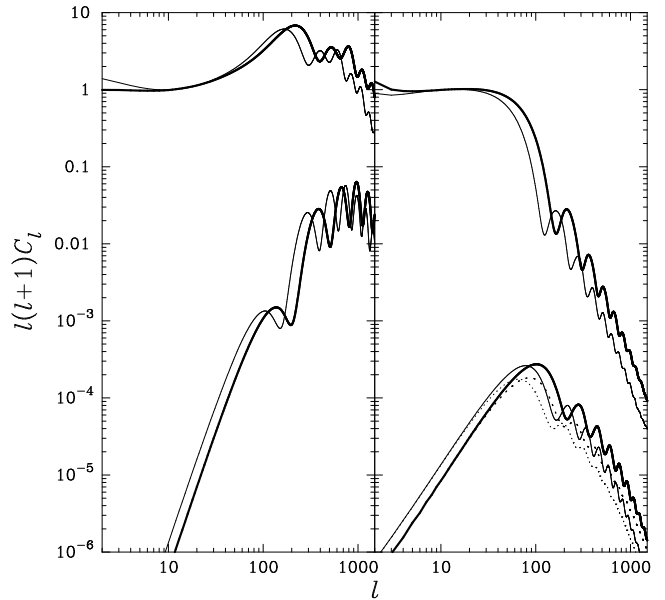
\figcaption[plot.eps]{Scalar (left) and
tensor (right) intensity and polarization power spectra in a closed CDM model
($\Omtot=1.2$, $\Oml=0.8$; thin lines), and a flat model
($\Omtot=1$, $\Oml=0.6$; thick lines). In both cases
$\Omm=0.4$, $\Omb=0.045$, and
$H_0=65 \mathrm{kms}^{-1}\mathrm{Mpc}^{-1}$. The upper solid lines are the
intensity, the lower ones the electric component of the polarization
$C_\ell^{\cle\cle}$, and the dashed lines the magnetic component
$C_\ell^{\clb\clb}$. (Note scalar modes produce no magnetic polarization.)
The scalar and tensor intensities are normalized to unity at $\ell=10$.
\label{Cls}}
\vspace{\baselineskip}

\section{CONCLUSION}

We have presented the first calculation of the CMB power spectra, including
the effects of polarization, in closed FRW models. We have implemented
the efficient line of sight algorithm for general geometries, using
covariantly defined, gauge-invariant variables, thus allowing accurate and
rapid modelling over the full volume of parameter space of FRW models.
Our Fortran 90 code, based on
\CMBFAST\ 
version 2.4.1,
is publically available at
http://www.mrao.cam.ac.uk/$\sim$aml1005/cmb.


\acknowledgements
A. Lewis is supported by a PPARC studentship, and A. Challinor by a Research
Fellowship at Queens' College, Cambridge.  A. Lasenby thanks the Royal Society
and Leverhulme Trust for support. We thank Uro\v{s} Seljak and Matias Zaldarriaga for making \CMBFAST\ publically available, and for suggesting we compare our results in closed
models against preliminary results from a pre-release version of
\CMBFAST\ which they have developed for closed models. We also thank
Arthur Kosowsky for making his WKB code available to us.



\end{document}